\documentclass[superscriptaddress,twocolumn]{revtex4}

\usepackage{amssymb}
\usepackage{amsfonts}
\usepackage{amsmath}
\usepackage{graphicx}
\usepackage{dcolumn}
\usepackage{bm}
\usepackage[latin1]{inputenc}
\begin{document}
\title{Role of the remnant symmetries in gravitational theories based on absolute parallelism: a 2D standpoint}

\author{Franco Fiorini}
\email{francof@cab.cnea.gov.ar} \affiliation{Departamento de Ingenier\'{i}a en Telecomunicaciones, Consejo Nacional de Investigaciones Cient\'{i}ficas y T\'{e}cnicas (CONICET) and Instituto Balseiro (UNCUYO), Centro At\'{o}mico Bariloche, Av. Ezequiel Bustillo 9500, CP8400, S. C. de Bariloche, R\'{i}o Negro, Argentina.}
\author{Andronikos Paliathanasis}
\email{anpaliat@phys.uoa.gr}
\affiliation{Institute of Systems Science, Durban University of Technology, PO Box 1334,
Durban 4000, South Africa.}
\affiliation{Instituto de Ciencias F\'{\i}sicas y Matem\'{a}ticas, Universidad Austral de
Chile, Valdivia 5090000, Chile.}

\begin{abstract}
By using simplified 2D gravitational, non-Lorentz invariant actions constructed from the torsion tensor, we discuss the physical meaning of the remnant symmetries associated with the near-horizon (Milne) geometry experienced by a radial observer in Schwarzschild spacetime. We then fully characterize the remnant symmetries corresponding to this near-horizon 2D geometry by solving the motion equations adapted to 2D Milne space. This symmetries, which represent special or privileged \emph{diads}, acquire the form of uniformly accelerated (Rindler) observers whose constant acceleration is proportional to the black hole mass $M$.
\end{abstract}
\maketitle

\section{Introduction}

Perhaps the most profound open problem concerning $f(T)$ gravity and its fertile progeny is the concise counting of the dynamical degrees of freedom (DoF), as well as the physical impact they would have in regard to the description of the gravitational field. This is actually an undergoing, glowing research area in which no general consensus seems to exists at the moment. Since its introduction circa fifteen years ago \cite{Nos3},\cite{BYF}, $f(T)$ gravity has given birth to an important flow of studies (see Ref. \cite{review} for an up to date review), but only in recent years the attention was paid to more formal, conceptual and fundamental aspects within the theory. After the first inkling in relation with the number of extra DoF \cite{Miao}, the bulk of the discussion on this subject started to develop recently \cite{RafaMajo1}, \cite{Nester}, not without some disparity in the results.

Another fundamental aspect closely related to the investigation of the DoF, is the breaking of the Lorentz invariance. The \emph{covariant} formulation started in \cite{Krsak1} (see also \cite{Krsak3}) is, in certain sense, opposed in nature to the original \emph{pure tetrad} approach and its consequent remnant symmetries, presented in \cite{Nos4}. So it is as one would seem to be in the position of having to choose between obtaining solutions at the expense of introducing a rather gloomy spin connection whose dynamics is not governed by genuine motion equations (former), or to be intuitive enough as to conceive intricate, not always well motivated ansatze in order to arrive to a set of consistent dynamical equations for the \emph{entire} tetrad field (latter). In both cases, the number -and specially- the \emph{physical nature} of the additional DoF seems to be uncannily hard to unmask. For a critical discussion concerning both approaches, we refer the reader to \cite{Nos5}-\cite{Manu}.

What seems to hamper the physical interpretation of the emergent DoF in $f(T)$ gravity is, in our opinion, the well funded reflect action of thinking in terms of the metric field $\textbf{g}$ alone, as a counterpart of having in mind the entire tetrad field $e^a$; clumsily, we are often lean to think that the \emph{only} influence of the extra DoF will be at the level of the metric, embodying thus some sort of modified behavior concerning the causal geodesics of a given spacetime with metric $\textbf{g}=e^ae^b\eta_{ab}$.  But the tetrad field is a quite more primary object, and it represents a myriad of different observers adapted to that \emph{same} geometry. In GR, where all is about the metric, the observers have no consequences whatsoever on the structure of the spacetime they observe. Nonetheless, under certain physical situations of interest, not all the tetrads representing observers are equally \emph{natural}. A Fermi-Walker transported tetrad, for instance, assures a succession of local non-rotating laboratories in Minkowski space, moving along a given world line with 4-velocity $v_{\mu}=e^0_{\mu}$. This could be certainly the most natural laboratory in arbitrary accelerated motion but, leaving aside an imminent dizziness, nothing prevents our laboratory from being a rotating one. Thorough discussions about the physical significance of the tetrad field as a reference frame adapted to a given observer were developed in the context of the Teleparallel Equivalent of General Relativity (TEGR), see \cite{mac3}-\cite{mac5}.

If we try to tip the balance in favor of the pure tetrad approach, as we do, we have to be ready to accept the presence of really privileged observers, at least in situations where the Lorentz invariance is strongly broken. Additionally, we have to be able to explain where this privilege comes from, and what these observers represent, both as a dynamical outcome of the theory. This article is a tentative to do so by constructing a couple of non local Lorentz invariant, torsion-based theories in the affable environment characterizing 2D gravity. These constructions are motivated by the near-horizon geometry experienced by a radial observer in Schwarzschild spacetime, which is described by a 2D Milne model as viewed from inside the black hole horizon. This simplification permits us to approach the main subject of this article, namely, the remnant symmetries underlying a given space; recognizing Milne space as the asymptotic state of the regular solution studied, we characterize its remnant group of symmetries and discuss the physical role they play in the description of the gravitational field near the horizon.

We mention that 2D Poincare gauge gravity was investigated before in \cite{revi0}. It was found there that the field equations are integrable while, specifically, black hole solution and propagating degrees of freedom exist. However, in this study we are interested in the role played by the remnant (or restricted) symmetries arising in concrete, 2D models which are not invariant under the local Lorentz group. Lower-dimensional gravitational theories, in general, constitute interesting toy-models that help to generate new ideas and to stimulate new insights into their higher dimensional counterparts. We refer the reader to the pedagogical review \cite{revi1}, and to the articles \cite{revi2,revi3} in what concerns to the significance of the Poincare group in 2D gravity.

\section{2D-Gravity built upon the torsion 2-form $T^{a}$}

\subsection{Near horizon geometry and Milne space}
Schwarzschild spacetime looks very different depending on what side of the horizon we are. In particular, from an interior point of view, it can be interpreted as a homogeneous and anisotropic cosmology with line element
\begin{align}
  ds^2 = dt^2-a^2(t)\,dx^2-b^2(t)\,d\Omega^2,
  \label{metrica}
\end{align}
where $a(t)$ and $b(t)$ are the scale factors depending only on the proper time and $d\Omega^2= d\theta^2+ \sin^2\negmedspace\theta d\phi^2$ is the line element of the 2-sphere. Kantowski and Sachs \cite{KS66} showed that metric (\ref{metrica}) represents a solution of the vacuum Einstein field equations if
\begin{align}
  a(t)=a_{1}\,\tan\left[\eta(t)\right],\quad
  b(t) = b_{1}\,\cos\negmedspace^2\left[\eta(t)\right],
  \label{func-a}
\end{align}
where the scale factors are written in terms of the time function $\eta(t)$ defined implicitly by
\begin{align}
 t-t_{0} =b_{1} (\eta+\sin\eta\cos\eta),
  \label{imptime}
\end{align}
that is,
\begin{equation}
dt=b_{1}\left[ 1+\left( \cos ^{2}\eta -\sin ^{2}\eta \right) \right] d\eta.
\end{equation}
 The integration constants $a_{1}$ and $b_{1}$ are constraint as $-\pi/2\leq a_{1}<0$ and $b_{1} \neq 0$. The constant $a_{1}$ is basically $2GM$ in units where $c=1$, see \cite{KS66}.

Let us inquire on how a radial observer experiences the near horizon geometry from an inside point of view. According to this cosmological interpretation, the black hole singularity corresponds to $\eta=\pi/2$, and the event horizon, in turn, to $\eta=0$. Thus, near the horizon we can invert (\ref{imptime}) and obtain $ \eta(t)\approx t/2 b_{1}$, after fixing $t_{0}=0$ without loss of generality. The scale factors result at the lowest order
\begin{align}
  a(t)\approx a_{1}\,t ,\quad
  b(t) \approx b_{1}\,(1-t^2/4\,b_{1}^2),
  \label{scalelow}
\end{align}
after redefining the integration constant $a_{1}$. Hence, constant $\theta,\phi$ observers will experience the near horizon geometry as if it were the one corresponding to a $D=2$ metric
\begin{equation}\label{metint}
ds^2=dt^2-a_{1}^2\,t^2dx^2,
\end{equation}
which is nothing else than the line element of 2D Milne space. This simple result motivates us to study 2D cosmological, asymptotically Milne spacetimes of the form
\begin{equation}\label{metint}
ds^2=dt^2-a(t)^2dx^2,
\end{equation}
where $a(t)$ is a $\mathcal{C}^1$ function verifying $a(t)\rightarrow \alpha\, t$ as $t\rightarrow \pm\infty$ for some constant $\alpha$. Whenever possible, we will try to assess the possibility of having regular solutions for which $a(t)\neq 0$ for any finite $t$. This condition assures the geodesic completeness of the cosmological manifold under consideration, and then, of the dimensionally reduced black hole interior. This is not really necessary for the discussions to be developed in section \ref{thermo} below, but it might be useful for a further understanding of the recently studied regular solutions found within the context of 4D Born-Infeld-$f(T)$ model \cite{BF1}.

\subsection{Some toy models}

Bearing in mind our objective of constructing non-local Lorentz invariant schemes for 2D-gravity, we present here two models which modestly serve to our purposes.
The two models are inspired by four dimensional theories formulated on the basis of absolute parallelism, as $f(T)$ gravity and Born-Infeld determinantal gravity, and both of them are constructed upon the \emph{zweibein} or \emph{diad} $e^a_{\mu}$ and its first derivatives alone.

If we think on replacing the superpotential 2-form $S^{a}$ appearing in the definition of the Weitzenb\"{o}ck invariant $T=S^{a}T_{a}$, by the torsion 2-form itself, we can easily conceive a 2D version of $f(T)$ gravity (remember that $S^{a}$ is identically null in 2D). Then, 2D-$f(\mathbb{T})$ gravity is governed by the action
\begin{equation}\label{accionefe}
I_{f}=\frac{1}{2\kappa } \int\,e \,f(\mathbb{T})\,d^2x,
\end{equation}
where $\kappa$ is a coupling constant with units of squared length (see the comment below Eq. (\ref{defy9})). Here, the object $\mathbb{T}$ is
\begin{equation}\label{escalardew}
\mathbb{T}=T^{a}_{\,\,\mu\nu}T_{a}^{\,\,\mu\nu},
\end{equation}
being $T^{a}=de^a$ the torsion 2-form, which in local coordinates reads $T^{a}_{\,\,\mu\nu}=\partial_{\mu}e^a_{\,\,\nu}-\partial_{\nu}e^a_{\,\,\mu}$. The action integral (\ref{accionefe}) is manifestly non-invariant under local Lorentz
transformations due to the fact that the spin connection $\omega^a_{\,\,b}$ was deliberately zeroed. This deprives $T^{a}$ from being a genuine Lorentz vector even though it transform as a tensor under general coordinate transformations. The geometric object (\ref{escalardew}) might be interpreted as the ''2D-version of the Weitzenb\"{o}ck invariant $T=S^{a}T_{a}$", which is a scalar only under general coordinate transformations. This is the basis of the pure tetrad approach.

The motion equations coming from varying the action (\ref{accionefe}) with respect to the diad components $e^a_{\mu}$ are
\begin{align}
  -4\bigl[T^{\rho\mu a} T_{\rho\mu\nu}-e^{-1}\partial_\mu(e\ &T^{a\mu}_{\,\,\,\,\,\,\,\nu})]
  f^{\prime }-4\,T^{a\mu}_{\,\,\,\,\,\,\,\nu} \partial_\mu \mathbb{T}\, f^{\prime \prime }+\nonumber\\
   &+e_{\,\,\nu}^a f = 2\kappa\, T_{\,\,\nu}^{a},
  \label{ecuaciones}
\end{align}
where $f^{\prime}$ and $f^{\prime\prime}$ refers to derivatives with respect to $\mathbb{T}$, and $T_{\,\,\nu}^{a}$ is the diad-projected energy momentum tensor $T_{\,\,\nu}^{a}=e^a_{\,\,\mu}T^{\mu}_{\,\,\nu}$. These equations can be obtained from the 4D-$f(T)$ motion equations by the simple rule of replacing $S^{a}_{\,\,\,\mu\nu}$ by $T^{a}_{\,\,\,\mu\nu}$.

The prescription of dimensionally reducing the 4-world by taking $S^{a}_{\,\,\,\mu\nu}\rightarrow T^{a}_{\,\,\,\mu\nu}$ also permits us to construct a 2D version of the already know 4D Born-Infeld (BI) determinantal gravity of Refs. \cite{Nos1}, \cite{Nos2}. In this case we have \cite{Vatu}
\begin{equation}\label{ac2d}
I_{BI}=\frac{\lambda}{2\kappa} \int\, \Big[\sqrt{\mid g_{\mu\nu}+2\lambda^{-1}F_{\mu\nu}\mid}-\sqrt{\mid g_{\mu\nu}\mid}\Big]\,d^2x,
\end{equation}
where $\mid...\mid$ is a shorthand for $Abs(Det(...))$, and $\lambda$ is the BI constant.

Tensor $F_{\mu\nu}$ reads
\begin{equation}\label{efe}
F_{\mu\nu}= \alpha\, A_{\mu\nu}+\beta\, B_{\mu\nu}+\gamma\, C_{\mu\nu},
\end{equation}
where in 2D the different contribution are
\begin{eqnarray}\label{pieces}
A_{\mu\nu}&=&T_{\mu\sigma\rho}T_{\nu}^{\,\,\,\sigma\rho},\notag\\
B_{\mu\nu}&=&T_{\sigma\mu\rho}T^{\sigma\,\,\,\rho}_{\,\,\,\nu},\notag\\
C_{\mu\nu}&=&g_{\mu\nu}T^{\rho\sigma\tau}T_{\rho\sigma\tau}=g_{\mu\nu} \mathbb{T}.
\end{eqnarray}
Factoring out $e=\sqrt{\mid g_{\mu\nu}\mid}$ in (\ref{ac2d}), we see that the action integral can be written in the equivalent form
\begin{equation}\label{ac2d2}
I_{BI}=\frac{\lambda}{2\kappa} \int\,e \,\Big[\sqrt{\mid \mathbb{I}+2\lambda^{-1}\mathbb{F}\mid}-1\Big]\,d^2x,
\end{equation}
where $\mathbb{I}$ is the identity and $\mathbb{F}=F_{\mu}^{\,\,\nu}$. If needed, the radicand in (\ref{ac2d2}) can be explicitly computed as
\begin{equation}\label{detee}
\mid \mathbb{I}+2\lambda^{-1}\mathbb{F}\mid=1+\frac{2}{\lambda}\,Tr(\mathbb{F})+\frac{2}{\lambda^2}\,[Tr^2(\mathbb{F})-Tr(\mathbb{F}^2)].
\end{equation}
Determinantal actions for the gravitational field are almost as old as GR \cite{Edd}. They were rediscovered some time ago \cite{deser}, see Ref. \cite{Gonzalo} for a comprehensive account on the matter.

We are particularly interested in the low energy limit of this theory. If we expand the action (\ref{ac2d}) or (\ref{ac2d2}) in powers of the small quantity $\lambda^{-1}$, we obtain at the lowest order
\begin{equation}\label{accionlow}
I_{\downarrow}=\frac{1}{2\kappa} \int\,e \,\mathbb{T}\,d^2x.
\end{equation}
This is the closest we can get to a ``2D-TEGR". However, this is just a metaphor, because there is not such a thing in 2D. The action (\ref{accionlow}) is of course also obtainable from (\ref{accionefe}) by taking $f=Id$. The motion equations arising from (\ref{accionlow}) then are
\begin{equation}
    \mathbb{T}\,e^{a}_{\, \,\nu}  -4\, T^{\sigma\rho a}T_{\sigma\rho \nu}  +4\, e^{-1} \partial_{\mu} \left( e\, T^{a\mu}_{\,\,\,\,\,\, \,\nu} \right) = 2\kappa\, \, T^{a}_{\, \,\nu}.
\label{eq: ecTT}
\end{equation}
Hence, this equations describe low energy limit of the two different 2D gravitational schemes in question. They will be analyzed in detail in section \ref{thermo} in regard with the role played by the asymptotic remnant symmetries. As a reassuring historical remark, we mention that (\ref{accionlow}) represents the 2D-version of \emph{Einstein's unified field theory}, see \cite{Delp} for a compendium of translated papers on the matter.

\subsection{Solutions having Milne asymptotics}

As a working example, we shall pay attention now to the cosmological solutions of the theory (\ref{ac2d}). The equations of motion coming from (\ref{ac2d}) or (\ref{ac2d2}) after varying with respect to the diad components $e^a_{\mu}$, are somewhat complicated and rather tedious to deal with in general, see \cite{Vatu} for details. However, we only need them for a \emph{seed} diad of the form $e^a_{\mu}=diag(1,a(t))$ leading to the metric (\ref{metint}). We have opportunely discussed the suitability of these seeds frames in what concerns to the parallelization of 4D Friedmann-Robertson-Walker cosmological manifolds, so we refer the reader to \cite{Nos6} for a deeper discussion. Later on in the next section we will affect this diad with a local Lorentz boost which, of course, does not have any impact on the metric, but which proves to be crucial with regard to the interpretation of the remnant symmetries.

The $(t,t)$-motion (constraint) equation coming from the above diad is:
\begin{eqnarray}\label{valin}
\frac{1-\frac{A\,B\,H^4}{\lambda^2}}{\sqrt{(1-\frac{A\, H^2}{\lambda})(1-\frac{B\, H^2}{\lambda})}}-1=\frac{2\kappa}{\lambda} \rho,
\end{eqnarray}
where $A=-2(\beta+2\gamma)$ and $B=-2(2\alpha+\beta+2\gamma)$, for arbitrary $\alpha$, $\beta$ and $\gamma$.

Because of the automatic conservation of the energy-momentum tensor (which we assume having the perfect fluid form $T^\mu_\nu=diag(\rho,-p)$, $p=\omega\rho$ in the comoving frame and $\omega$ is a constant), we have
\begin{eqnarray}\label{energyden}
\rho(t)=\rho_{0}\Big(\frac{a_0}{a(t)}\Big)^{(1+\omega)},
\end{eqnarray}
where $\rho_{0}$ and $a_0$ are constants.

For a two-dimensional spacetime the $\omega=0$ corresponds to dust matter (as usual), and $\omega=1$ to a radiation fluid. In the latter, $\rho(t)\propto a(t)^{-2}$, which will be important in a moment. Finally, the scale factor $a(t)$ is determined thus by Eqs. (\ref{valin}) and (\ref{energyden}).

We have essentially two different cases of interest, namely:

\bigskip

1) Case $A=B$: Here the constraint Eq. (\ref{valin}) reads
\begin{equation}\label{valincaso1}
H^2\propto \rho(t)\,\,\Rightarrow \,\,H^2\propto a(t)^{-(1+\omega)},
\end{equation}
because of (\ref{energyden}). Let us notice that for a radiation fluid it follows $a(t)\propto|t|$. This is exactly -no just asymptotically- 2D-Milne spacetime. This ``low energy'' regime ($\lambda$ is absent here), then also appears as a pure state of the determinant action.

\bigskip

2) Case $A=0$ or $B=0$: In this case we find
\begin{equation}\label{valincaso2}
\frac{1}{\sqrt{1-\frac{H^2}{\lambda}}}-1=\frac{2\kappa}{\lambda}\, \rho(t),
\end{equation}
after reabsorbing $A$ or $B$ (depending on the case), into $\lambda$. Remarkably, this equation has the same structure than the one coming from the BI-$f(T)$ theory of Ref. \cite{Nos3}. It leads to exact regular solutions for all physical values of $\omega$. In particular, for $\omega=1$, the solution consist of an asymptotic de Sitter stage as $t\rightarrow -\infty$, and an asymptotic Milne one as $t\rightarrow \infty$ (expanding branch). Conversely, the contracting branch is characterized by a Milne asymptotics as $t\rightarrow -\infty$ and a de Sitter one as $t\rightarrow \infty$.  Let us briefly account for this.
We define the new variable
\begin{equation}\label{defy}
\texttt{y}= \frac{\lambda}{2\kappa\,  \rho(t)}=\frac{\lambda}{2\kappa\,  \rho_{0}} \Big(\frac{a(t)}{a_{0}}\Big)^{(1+\omega)},
\end{equation}
such that
\begin{equation}\label{defy2}
\frac{\dot{\texttt{y}}}{\texttt{y}}= (1+\omega)H.
\end{equation}
In this way, Eq. (\ref{valincaso2}) leads us to
\begin{equation}\label{defy3}
\dot{\texttt{y}}=\pm\mathcal{A}\,\frac{\texttt{y}}{1+\texttt{y}}\sqrt{1+2\texttt{y}},
\end{equation}
which, after integration, can be cast as
\begin{equation}\label{defy4}
\pm \mathcal{A}\,t\pm c=\sqrt{1+2\texttt{y}}+\ln\Big[\frac{-1+\sqrt{1+2\texttt{y}}}{1+\sqrt{1+2\texttt{y}}}\Big],
\end{equation}
where $\mathcal{A}=\sqrt{\lambda}\,(1+\omega)$ and $\pm c$ are integration constants. These are the exact (implicit) solutions for the scale factor, corresponding to contracting and expanding phases. In our view, they describe the radial motion in regular black and white holes interiors, respectively.

The asymptotic behavior of the scale factor is easily obtained by expanding in the variable $\texttt{y}$ the right hand side of Eq. (\ref{defy4}) in the corresponding limits. If $\texttt{y}<<1$ (this corresponds to $\rho>>1$ or, equivalently, $a(t)<<1$), Eq. (\ref{defy4}) results
\begin{equation}\label{defy5}
\pm \mathcal{A}\,t\pm c\approx  \ln(\texttt{y}/2),
\end{equation}
or, in terms of the scale factor,
\begin{equation}\label{defy6}
\exp(\pm \mathcal{A}\,t\pm c)\approx \frac{\lambda}{4\kappa \rho_{0}} \Big(\frac{a(t)}{a_{0}}\Big)^{(1+\omega)},
\end{equation}
then
\begin{equation}\label{defy7}
a(t)\approx \tilde{a}_{0} \exp(\pm\sqrt{\lambda}\,t\pm c),
\end{equation}
where $\tilde{a}_{0}=a_{0}(\lambda/4\kappa\rho_{0})^{(1+\omega)}$ and $c$ was redefined.

We observe that for $t\rightarrow\pm \infty$ (depending if we are in the contracting or expanding branch), the scale factor corresponds to a de Sitter stage. This will render the black (or white) hole geodesically complete for any physical $\omega$.

In turn, if we consider $\texttt{y}>>1$, Eq. (\ref{defy4}) reads
\begin{equation}\label{defy8}
\pm \mathcal{A}\,t\pm c\approx \sqrt{2\texttt{y}},
\end{equation}
that is,
\begin{equation}\label{defy9}
\Big(\sqrt{\kappa} (1+\omega) \, t\pm c\Big)^2 \approx \Big(\frac{a(t)}{a_{0}}\Big)^{(1+\omega)}.
\end{equation}

After a redefinition of $c$, which hereafter we will zeroed without loss of generality. Milne spacetime $a(t)\propto |t|$ emerges asymptotically for a radiation dominated universe ($\omega=1$). With the help of (\ref{metint}) and (\ref{defy9}) we can mention that the units of $\kappa$ are just $[length]^2$, because $a_{1}^2=4G^2M^2$.

\begin{figure}[tbp]
\centering\includegraphics[width=0.45\textwidth]{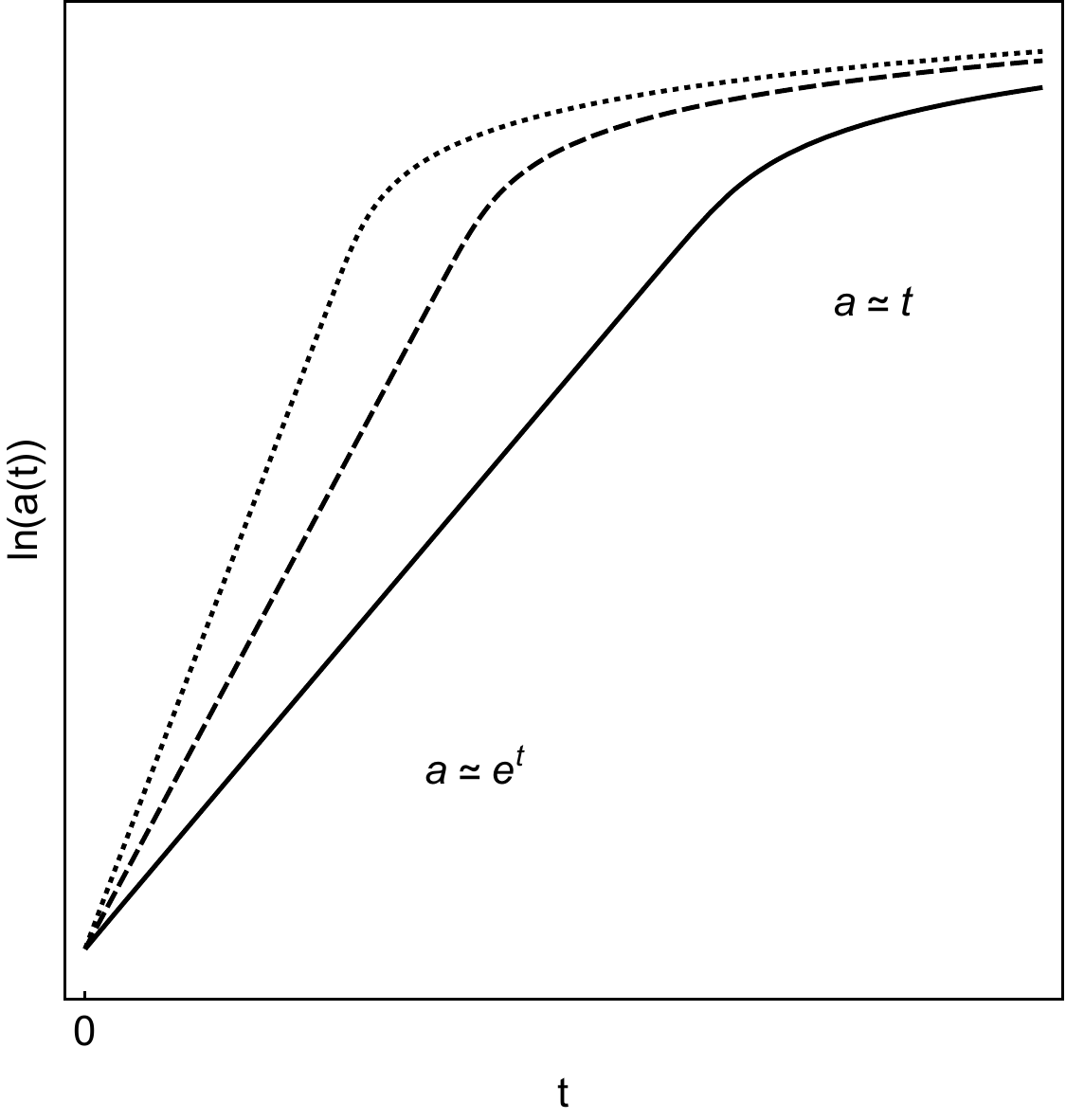}
\caption{Evolution of the scale factor $\ln a\left( t\right) $
as provided by the equation (\protect\ref{defy4}), in the expanding branch of solutions. Plot is
for $\protect\omega =1$ and solid line is for $\lambda=1/5$, dashed line is for $\lambda=1/2$ and dotted line is for $\lambda=1$. We observe
that for small values of $t$, the behaviour of the scale factor is
exponential, while for large values of $t$, the Milne universe is recovered.}
\label{fig1}
\end{figure}

Notice that $\lambda$ is absent in this solution. This reinforces the fact that Milne is a sort of low energy regime of the theory, consistent with the fact that it duly describes the spacetime near the horizon. This is the kind of behavior we are looking for. In Fig. \ref{fig1} we present the evolution of the scale factor for the expanding branch as it is provided by the equation (\ref{defy4}), from where
we observe the asymptotics just discussed.

In order to complete the understanding of the solution we now derive the Ricci scalar for the two-dimensional spacetime in discussion. For the metric (\ref{metint}) we have $R=2\ddot{a}/a$, or $R=2\left(\dot{H}+H^{2}\right)$. Due to the fact that any two-dimensional space is conformally flat, the Riemann curvature tensor $R^{\mu}_{\,\,\nu\rho\sigma}$ and the Ricci tensor $R_{\mu\nu}$ can be written in terms of the scalar $R$, so that $R$ alone completely determines the local geometry. Moreover, when $R=const.$, the two dimensional space is maximally symmetric. Indeed for $R=0$, the spacetime is
flat, which is possible when $a\left( t\right)$ is linear in time, that is, on the Milne solution, and when $a\left( t\right) =const.$ which corresponds to Minkowski space. On the other hand, when $R$ is a non-null constant the scale factor is exponential, which corresponds to de
Sitter solution. Furthermore, when $a\left( t\right) $ is expressed by a power-law function, $R$ is asymptotically zero for large
values of $t$.

The Ricci scalar $R$ associated with the solution (\ref{defy4}) can be explicitly obtained in terms of the variable $\texttt{y}$ defined in (\ref{defy}), with the help of (\ref{defy2}) and (\ref{defy3}). It reads
\begin{equation}\label{elR}
R=2\lambda\left[\frac{1+3\texttt{y}+(1-\omega)\texttt{y}^2}{(1+\texttt{y})^3}\right]
\end{equation}
Thus for any $\omega$, we observe that for small values of $\texttt{y}$ (or, equivalently, $a(t)$), $R\rightarrow 2\lambda $, while for large values of the scale factor, $R\rightarrow 0$. Hence, the spacetime (\ref{defy4}) dynamically describes the transition between two constant curvature asymptotic spaces.

Of course, $R$ is incapable of highlighting the global differences existing between, for instance, Minkowski and Milne spaces. What we need is to ask about this to the diad field instead. This is what next two sections are devoted to.

\section{Remnant symmetries and accelerated observers}\label{thermo}

We have already mentioned that the $\lambda\rightarrow\infty$ limit of the action integral (\ref{ac2d2}) is given by expression (\ref{accionlow}). So, let us study under what circumstances Milne universe is a solution of the low energy theory. It is our intention to explore the remnant group associated to that solution. This is relevant because in the regular solution (\ref{defy4}), Milne emerges as the asymptotic phase of the evolution, then, to characterize its remnant group will allow us to interpret the nature of the \emph{asymptotic group of remnant symmetries}.

In general, the pedestrian way of looking for the remnant symmetries of a given space characterized by the scale factor $a(t)$, is to locally boost the seed-diad $e^a_{\mu}=diag(1,a(t))$, and to inquire then under what circumstances the motion equations remain unchanged. In a general situation, seeds diads (or tetrads in 4D) are essentially the simplest ansatze consistent with the motion equations. By no means they are diagonal nor natural in general, see Ref. \cite{Nos6} in regard to the seed tetrads associated with 4D, $K=1,-1$ FRW spacetimes. However, spatially flat FRW spaces are correctly described by the diagonal tetrad, which constitutes a parallelization of the manifold in question. This tetrad leads to a consistent set of motion equations for the scale factor $a(t)$, for any function $f(T)$. See \cite{Nos6} for details on this point.

In the present circumstances, after boosting the diad $e^a_{\mu}=diag(1,a(t))$ we obtain $e^{b}=\Lambda^b_{\,\,a}(t,x)\,e^{a}$, or
\begin{eqnarray}\label{diadboost}
e^{1}&=\cosh[\phi(t,x)]\,dt+a(t)\sinh[\phi(t,x)]\,dx\notag\\
e^{2}&=\sinh[\phi(t,x)]\,dt+a(t)\cosh[\phi(t,x)]\,dx,
\end{eqnarray}
where the explicit dependence of the \emph{booston} $\phi(t,x)$ on the space-time coordinates was emphasized.

Consider now the equations of motion (\ref{eq: ecTT}) for the diad (\ref{diadboost}). They can be combined to get (from now on, dots refer to time derivatives, and primes to derivatives with respect to $x$)
\begin{eqnarray}
\phi''-a\, \phi'\, \dot{\phi}&=&0\label{ecsbaja1}\\
\ddot{\phi}+\dot{\phi}\Big(H-\frac{\phi'}{a}\Big)&=&0\\\label{ecsbaja2}
H^2-\dot{\phi}^2+2\frac{\dot{\phi}'}{a}-\frac{\phi'^{\,2}}{a^2}&=&\kappa \rho\\\label{ecsbaja3}
H^2+\dot{H}&=&\kappa(\rho-p)/2
\label{ecsbaja4}
\end{eqnarray}
Just as in GR, not all the equations are independent. In the case of a global ($\phi=constant$) symmetry, we see that the equations are remarkably similar to the ones of GR:
\begin{eqnarray}
H^2&=&\kappa \rho\label{ecsgr1}\\
H^2+\dot{H}&=&\kappa(\rho-p)/2.\label{ecsgr2}
\end{eqnarray}
By differentiating Eq. (\ref{ecsgr1}) with respect to time and combining it with (\ref{ecsgr2}), we obtain the conservation equation
\begin{equation}\label{conservacion}
\dot{\rho}+H(\rho+p)=0.
\end{equation}

Being the system (\ref{ecsbaja1})-(\ref{ecsbaja4}) representative of the low energy regime of the BI motion equations, we are interested in the Milne solution. This spacetime arises in two different context, namely, as a solution in the presence of a radiation fluid, and as a vacuum solution as well; however, the interpretation of the remnant symmetries involved is different depending on the case. Let work this out in detail.

If we set $a(t)=a_{1}t$, and $p=\rho$, the system (\ref{ecsbaja1})-(\ref{ecsbaja4}) becomes
\begin{eqnarray}
\phi''-a_{1}t\, \phi'\, \dot{\phi}&=&0\label{ecsbaja11}\\
\ddot{\phi}+t^{-1}\dot{\phi}\,(1-a_{1}^{-1}\phi')&=&0\label{ecsbaja21}\\
t^{-2}(1-a_{1}^{-2}\phi'^{\,2})+2a_{1}^{-1}t^{-1}\dot{\phi}'-\dot{\phi}^{2}&=&b\, t^{-2},\label{ecsbaja31}
\end{eqnarray}
where $b=\kappa a_{1}^{-2} \rho_{0}\, a_{0}^2$ and  Eq. (\ref{ecsbaja4}) is automatically verified. Hence, $b=0$ captures the situation in which no matter fields are present ($\rho_{0}=0$).

What local symmetries are admissible in regard with Milne space? If we have $\phi=\phi(x)$, then Eq. (\ref{ecsbaja21}) is solved at once, and the other two lead us to
\begin{equation}\label{boostonx}
  \phi(x)=\phi_{0}\,x+\phi_{1},\,\,\,\,\phi_{0}^2/a_{1}^2=1-b,
\end{equation}
for arbitrary $\phi_{1}$. If, in turn, $\phi=\phi(t)$, Eq. (\ref{ecsbaja11}) is automatically fulfilled and (\ref{ecsbaja21})-(\ref{ecsbaja31}) are solved by the booston
\begin{equation}\label{boostont}
  \phi(t)=\phi_{0}\,\ln t+\phi_{1},\,\,\,\,\phi_{0}^2=1-b,
\end{equation}
again for arbitrary $\phi_{1}$. In both (\ref{boostonx}) and (\ref{boostont}) the choice $\phi_{0}=0$ conducts us to the global, trivial booston. After this observation we can safely take from now on $\phi_{1}=0$, $\phi_{0}\neq0$ in (\ref{boostonx}) and (\ref{boostont}).

If we move away from boostons depending solely on one coordinate and define the new variable $z=\frac{a_{1}x-\ln t}{a_{1}x+\ln t}$, we determine a third solution of (\ref{ecsbaja11})-(\ref{ecsbaja31}). In this case we have $\phi =\phi \left( z\right) $ of the form
\begin{equation}\label{boostontx}
\phi \left( z\right) =\ln z+z_{0},\,\,\,\, b=1\text{.}
\end{equation}

\bigskip

We now proceed to interpret the physical meaning of the symmetries just obtained, starting from (\ref{boostonx}), which presents itself as the most laborious. Let us make a three-step simple coordinate change without performing any further Lorentz transformation, according to the following scheme:

a) Let us take
\begin{equation}
   a_{1}t=\sqrt{2a_{1}\bar{T}-1},
\label{change1}
\end{equation}
where, clearly, $\bar{T}>1/2a_{1}$, so we are considering positive values of $t$, which corresponds to the positive or expanding branch of Milne Universe (similar results will hold in the negative, contracting branch). Remember that in Milne space we have $x\in \mathbb{R}$ and $t\in \mathbb{R}-\{0\}$ in order to avoid the coordinate singularity at $t=0$. However, all that matters lies in the limit $t\rightarrow \infty $ (in the expanding branch under consideration), where BI gravity coincides with the low energy theory in question.
After this coordinate change the diad \ref{diadboost} becomes
\begin{eqnarray}\label{diadboostchange1}
e^{1}&=\frac{\cosh[\phi_{0}\,x]}{\sqrt{2a_{1}\bar{T}-1}}\,d\bar{T}+\sqrt{2a_{1}\bar{T}-1}\,\sinh[\phi_{0}\,x]\,dx\notag\\
e^{2}&=\frac{\sinh[\phi_{0}\,x]}{\sqrt{2a_{1}\bar{T}-1}}\,d\bar{T}+\sqrt{2a_{1}\bar{T}-1}\,\cosh[\phi_{0}\,x]\,dx,
\end{eqnarray}
while the metric is
\begin{equation}
   ds^2=-\frac{d\bar{T}^2}{(2a_{1}\bar{T}-1)}+(2a_{1}\bar{T}-1)\,dx^2.
\label{metboostchange1}
\end{equation}

b) In the region $\bar{T}<1/2a_{1}$ the causal character of the coordinates $\bar{T}$ and $x$ is inverted. This is similar to what happen when crossing the Schwarzschild horizon $r=2M$, where the coordinate $r$ becomes timelike inside the horizon, while $t$ switches to spacelike. So, let us consider the change to the new coordinates $(T,\bar{x})$ defined as $\bar{x}=\bar{T}$ and $T=x$. The diad looks then
\begin{eqnarray}\label{diadboostchange2}
e^{1}&=\frac{\cosh[\phi_{0}\,T]}{\sqrt{1-2a_{1}\bar{x}}}\,d\bar{x}+\sqrt{1-2a_{1}\bar{x}}\,\sinh[\phi_{0}\,T]\,dT\notag\\
e^{2}&=\frac{\sinh[\phi_{0}\,T]}{\sqrt{1-2a_{1}\bar{x}}}\,d\bar{x}+\sqrt{1-2a_{1}\bar{x}}\,\cosh[\phi_{0}\,T]\,dT,
\end{eqnarray}
and the metric
\begin{equation}
   ds^2=- (1-2a_{1}\bar{x})\,dT^2+\frac{d\bar{x}^2}{(1-2a_{1}\bar{x})}.
\label{metboostchange2}
\end{equation}

c) The final step consists on performing yet another change to a new coordinate $X$ defined by
\begin{equation}
   1-a_{1}X=\sqrt{1-2a_{1}\bar{x}},
\label{change3}
\end{equation}
where obviously $X<a_{1}^{-1}$. It leads us to the final form of the diad
\begin{eqnarray}\label{diadboostchange3}
e^{1}&=\cosh[\phi_{0}\,T]\,dX+(1-a_{1}X)\,\sinh[\phi_{0}\,T]\,dT\notag\\
e^{2}&=\sinh[\phi_{0}\,T]\,dX+(1-a_{1}X)\,\cosh[\phi_{0}\,T]\,dT,
\end{eqnarray}
and the metric
\begin{equation}
   ds^2=- (1-a_{1}X)^2\,dT^2+dX^2.
\label{metboostchange3}
\end{equation}
This is a nice result, though not totally unexpected. Metric (\ref{metboostchange3}) corresponds to Rindler spacetime, i.e., the metric experienced by a uniformly accelerated observer in Minkowski space, having constant acceleration $a_{1}$. This is not surprising because we know that the near horizon Schwarzschild geometry in 4D can be decomposed into a direct product of 2D Rindler spacetime and a 2-sphere of constant radius, but we have omitted the 2-sphere from the beginning. In other words, the 2D near horizon geometry can be viewed as a Rindler space from the outside, or as a Milne space from the inside, a result very well understood and worked out, for instance, in Ref. \cite{Culetu1}.

What is really more important is the physical interpretation of the diad (\ref{diadboostchange3}). Of course, (\ref{diadboostchange3}) leads to the metric (\ref{metboostchange3}) for \emph{any} value of $\phi_{0}$. However, in order to interpret the diad (\ref{diadboostchange3}) as the one corresponding to an uniformly accelerated observer in flat space with acceleration $a_{1}$, we need to fix $\phi_{0}=a_{1}$; indeed, by virtue of the fact that we are in flat space, the diad (\ref{diadboostchange3}) can simply be viewed as a coordinate change of the Minkowski metric $\eta_{\mu\nu}=e^a_{\mu}e^{b}_{\nu}\eta_{ab}$, i.e. $e^a_{\mu}=\partial x^{a}/\partial x^{\mu}$, see Ref. \cite{gravitation}. But $\phi_{0}=a_{1}$ is precisely the vacuum case, where $b=0$, see Eq. (\ref{boostonx}). So, the remnant symmetry in this case is exactly the one needed to convert the solution near the horizon into a frame corresponding to a uniformly accelerated observer with acceleration $a_{1}$ \cite{frames1}. Take note that the key point leading to this identification is that the function $\phi_{0}\,T$ appearing in (\ref{diadboostchange3}) was not subjected to any coordinate transformation, except for the trivial switch $T=x$ needed to identify Milne and Rindler spaces in either side of the horizon (see point (b) above).

What can we say about the remnant symmetry displayed at (\ref{boostont})? The coordinate change $t=\exp(a_{1}\tau)$ brings the metric into the form
\begin{align}
ds^2=a_{1}^2\,\exp(2a_{1}\tau)[d\tau^2-dx^2]=a_{1}^2\,ds^2_{Rin},
  \label{confrin}
\end{align}
which is conformal to the line element corresponding to Rindler space written in conformal coordinates. Take note, however, that the conformal factor is constant, so it does not change at all neither the geometrical nor the topological aspects of the null-curvature Rindler space, but merely redefine the acceleration of the Rindler observer to a unitary value. We can see this by simply reescaling (\ref{confrin}) by means of $\eta=a_{1}\tau$ and $a_{1}x=X$ so as to obtain
\begin{align}
ds^2=\exp(2\eta)[d\eta^2-dX^2].
  \label{confrin2}
\end{align}
Meanwhile, under the same coordinate changes, the diad ends up being
\begin{eqnarray}\label{diadboostconf}
e^{1}&=e^{\eta}\Big[\cosh[\phi_{0}\,\eta]\,d\eta+\sinh[\phi_{0}\,\eta]\,dX\Big]\notag\\
e^{2}&=e^{\eta}\Big[\sinh[\phi_{0}\,\eta]\,d\eta+\cosh[\phi_{0}\,\eta]\,dX\Big],
\end{eqnarray}
which, again, embodies the diad adapted to a uniformly accelerated observer (with unit acceleration in this case), \emph{only} if $\phi_{0}=1$, corresponding to the vacuum case $b=0$ (see Eq. (\ref{boostont})). Due to the specific conformal coordinates in use, the diad (\ref{diadboostconf}) is also written in conformal coordinates. So, just as we did with the $x$-dependent booston (\ref{boostonx}), we can interpret the symmetry (\ref{boostont}) as being responsible for endowing the spacetime with observers in uniformly accelerated motion.

Lastly, with respect to the spacetime symmetry (\ref{boostontx}), we have much less to say, except for a few general remarks. Firstly, this booston exists only for $b=1$, which means that $a_{1}^2=\kappa \rho_{0} a_{0}^2$; then, if $a_{1}$ could allegedly be interpreted as an acceleration, it will depend on the matter content through $\rho_{0}$. Secondly, the absence of a multiplicative integration constant in front of the $\ln z$ term in (\ref{boostontx}), makes it topologically disconnected from the global symmetry associated with the constant booston. This means that this symmetry reveals itself as a detached state, disconnected from (\ref{boostonx}) and (\ref{boostont}), rendering the acceleration-like interpretation implausible. Undoubtedly, we need more work to clarify this point.

\section{Final comments}\label{further}

Sometimes, the local remnant symmetries associated with a given space can be anticipated at the level of the action. That is the case, for instance, of the local symmetries (\ref{boostonx}) and (\ref{boostont}) in the vacuum case $b=0$. As a matter of fact, the scalar $\mathbb{T}$ for the general diad (\ref{diadboost}) results
\begin{align}
\mathbb{T}=\frac{2(\dot{a}-\phi')^2}{a^2}-2\dot{\phi}^{2},
  \label{eltt}
\end{align}
which is identically null for $a(t)=a_{1} t$ and the boostons (\ref{boostonx}) and (\ref{boostont}) with $b=0$, depending on the case. Then, despite the cosmological semblance associated with Milne solution, we see that the boostons in question lead to $\mathbb{T}=0$, canceling the (linear) time dependence of the Milne scale factor. In this case the vanishing of $\mathbb{T}$ captures the fact that we are really dealing with flat spacetime, notwithstanding the presence of a non trivial global geometry due to the existence of a horizon.

The fact that Milne space is just a portion of Minkowski space, leads us to ask what is the intrinsic difference between the two at the level of the diad field. The system (\ref{ecsbaja1})-(\ref{ecsbaja4}) adapted to Minkowski space, i.e., when we consider a constant scale factor (which we can safely set equal to one), and $b=0$, is
\begin{eqnarray}
\phi''-\phi'\, \dot{\phi}&=&0\label{ecmin1}\\
\ddot{\phi}-\phi'\,\dot{\phi}&=&0\label{ecmin2}\\
-\dot{\phi}^2+2\dot{\phi}'-\phi'^{\,2}&=&0\label{ecmin3}.
\label{ecsbaja}
\end{eqnarray}
Eqs. (\ref{ecmin1}) and (\ref{ecmin2}) imply $\square^2\phi=0$, which is solved in null coordinates $U=t-x$, $V=t+x$, by
\begin{equation}
\phi=\phi_{U}+\phi_{V},
\label{ondas}
\end{equation}
where $\phi_{U}\equiv\phi_{U}(U)$, $\phi_{V}\equiv\phi_{V}(V)$ are arbitrary (twice differentiable) functions. After plugging (\ref{ondas}) in (\ref{ecmin3}) we find
\begin{equation}
\phi_{U}=\log(u_{0}\,U+u_{1}),\,\,\, \phi_{V}=\log(v_{0}\,V+v_{1}),
\label{ondassol}
\end{equation}
which automatically solve (\ref{ecmin1}) and (\ref{ecmin2}) separately for arbitrary constants $u_{i},v_{i}$, $i=0,1$. Minkowski spacetime is then not only characterized by the metric $\eta_{ab}=diag(1,-1)$, but also by a set of observers given by (\ref{ondassol}). These observers have a rapidity $\textbf{w}$ of the form
\begin{equation}\label{boostmin}
\textbf{w}=\tanh(\phi)=\frac{-1+f^2(U,V)}{1+f^2(U,V)},
\end{equation}
where
\begin{equation}\label{boostmin2}
f^2(U,V)=(u_{0}\,U+u_{1})^2(v_{0}\,V+v_{1})^2.
\end{equation}
According to this view, to take $\phi_{0}=0$ in (\ref{boostonx}) and (\ref{boostont}), as well as $u_{0}=v_{0}=0$ in (\ref{ondassol}) (constant boostons), constitutes a sort of oversimplification which deprive us from knowing global aspects of the spacetime. Precisely, global boostons represent observers having constant speed and, by use them, we are unable to make a distinction between Milne and Minkowski spaces. We mention that still there is much work to do in relation to ascertain what really Minkowski spacetime is in the context of $f(T)$ gravity, see \cite{Golo2} and \cite{Golo3}.

Let us take note that now (\ref{eltt}), for $a(t)=1$ and the boostons (\ref{ondassol}), acquires the form of a massless scalar field lagrangian $\mathbb{T}=-2\eta^{\mu\nu}\phi_{,\,\mu}\phi_{,\,\nu}$, which vanishes for every well behaved function $\phi$ of the null coordinates $U$ and $V$; but the motion equations further restrict the function $\phi$ to be as in (\ref{ondassol}). This means that the procedure of making $\mathbb{T}=0$ in order to anticipate a symmetry constitutes just a convenient guiding principle from which we cannot expect much, not even in flat space. One has to be aware that this process might have serious limitations if one eagerly freezes \emph{too many} DoF at the level of the action.

Finally, let us notice that 2D spacetime (\ref{metboostchange3}) has a characteristic (Unruh) temperature $T_{U}=a_{1}/2\pi$ (in units where $k_{B}=\hbar=1$), associated with the existence of a (Rindler) horizon at $X=a_{1}^{-1}$. This can be seen by going to the Euclidean section and then removing the conical singularity by imposing periodicity in imaginary time, see, e.g. Ref. \cite{BHT} for a similar result in 2D. The fact that the remnant symmetry behind the diad (\ref{diadboostchange3}) corresponding to the vacuum case $\phi_{0}=a_{1}$ gives us $\phi_{0}=2 \pi T_{U}$, appears to be a pleasantly suspicious coincidence, perhaps associated with some topological invariant related to the diad field. This is certainly a stimulating suggestion which we are going to take care of in future researches.

Before we conclude it is important to mention that the 2D Milne universe can be the limit of other exact solutions of $f(T)$ gravity. Indeed, in the case of homogeneous and anisotropic spaces, within $f(T)$ gravity Kasner-like exact solutions exist \cite{kas1}. These exact solutions are generalizations of the Kanser solution where the powers of the scale factor for the background space take values on a non-unitary sphere, which means that the 2D Milne universe can be recovered. For instance, the Mixmaster universe in the limit where the spatial curvature is neglected, can be described asymptotically by Kasner-like solutions. Thus the results of this study may be applied also to get an insight in the behavior of the cosmological singularity.

We do not know at the moment if the conclusions here exposed may or may not be translated into the real 4D world, but we expect them to help us in following the scent towards a deeper understanding of the role played by the additional DoF in $f(T)$ gravity.

\bigskip

\emph{Acknowledgements.} The authors are indebted to Rafael Ferraro for his comments on this manuscript. FF is a member of Carrera del Investigador Cient\'{i}fico (CONICET), and his work is supported by CONICET and Instituto Balseiro (UNCUYO).

\end{document}